\documentstyle[12pt,a4,epsf]{article}

\newcommand{\be}{\begin{equation}}
\newcommand{\ee}{\end{equation}}
\newcommand{\ba}{\begin{eqnarray}}
\newcommand{\ea}{\end{eqnarray}}
\newcommand{\dis}{\displaystyle}

\begin{document}
\begin{titlepage}
\begin{flushright}
{UG-FT-103/99}\\
{hep-ph/9909245}\\
\end{flushright}
\vspace{2cm}
\begin{center}

{\large\bf The $\Delta I=1/2$ Rule for Kaons\footnote{Work supported
in part by CICYT, Spain (Grant No. AEN-96/1672)
and by Junta de Andaluc\'{\i}a, (Grant No. FQM-101),
and by the European Union TMR Network $EURODAPHNE$ (Contract
No. ERBFMX-CT98-0169).
Invited talk at ``High Energy Euroconference on Quantum Chromodynamics
 (QCD '99)'', 7-13 July 1999, Montpellier, France.}}\\
\vfill
{\bf Joaquim Prades}\\[0.5cm]
Departamento de
 F\'{\i}sica Te\'orica y del Cosmos, Universidad de Granada\\
Campus de Fuente Nueva, E-18002 Granada, Spain.\\[0.5cm]
\end{center}
\vfill
\begin{abstract}
We report on recent advances at understanding the $\Delta I= 1/2$
 rule for kaons.
We get  reasonable matching between short-- and long--distances
for scales between between 0.6 and 1.0 GeV and reproduce the
$\Delta I=1/2$ rule huge enhancement in the chiral limit.
A detailed analysis of the different contributions to the
relevant octet and 27-plet couplings is done.
For the $B_6^{(1/2)}( \mu) \equiv  \langle (\pi\pi)_0 | Q_6 | K \rangle
/ \left[\langle (\pi\pi)_0 | Q_6 | K \rangle {\Large|}_{N_c}\right]$
parameter, we get in the chiral limit $B_6^{(1/2)}(\mu) = 2.2 \pm 0.5$ 
for scales $\mu \in [0.6, 1.0]$ GeV.
\end{abstract}
\vfill
August 1999
\end{titlepage}

\section{Introduction}
Understanding the $\Delta I=1/2$ rule for kaons within QCD
has been a continuous challenge, see \cite{EdR} for a review.
Here, we report on recent work and advances at understanding
this empirical rule in the chiral limit \cite{BP99a}.
Due to the lack of space, we would like to concentrate on two main issues. 
The first one is the heavy $X_i$--bosons technique and the 
the matching procedure.
The second one is 
the anatomy of the different contributions to the octet coupling
enhancement. We also give results on the penguin operator $Q_6$ which are
relevant for $\varepsilon' / \varepsilon$. 

Within the Standard Model (SM), $K\to\pi\pi$ decay amplitudes can be
decomposed into definite isospin $0$ and $2$  amplitudes as follows
$[A\equiv-iT]$, 
\ba
A[K_S\to\pi^0\pi^0] &\equiv&
 \sqrt{\frac{2}{3}} A_0 - \frac{2}{\sqrt 3} A_2 \, , 
\nonumber \\
A[K_S\to\pi^+\pi^-] &\equiv&
 \sqrt{\frac{2}{3}} A_0 + \frac{1}{\sqrt 3} A_2 \, , 
\nonumber \\
A[K^+\to\pi^+\pi^0] &\equiv&  \frac{\sqrt 3}{2} A_2  \, .
\ea
Where we have included the final state interaction phases
$\delta_0$ and $\delta_2$ into the amplitudes $A_0$ and $A_2$
as follows
\ba
A_{0(2)} \equiv -i a_{0(2)} e^{i\delta_{0(2)}} \, . 
\ea
Performing a fit to experimental data on $K\to\pi \pi$
and $K\to \pi\pi\pi$ up to Chiral Perturbation Theory
(CHPT) order $p^4$, in ref. \cite{KMW91} obtained
\ba
\label{deltaI}
\left| \frac{A_0}{A_2} \right|^{(2)} = 16.4 \, 
\ea
to lowest order.
Unfortunately no fit uncertainties were quoted.
This is the so--called $\Delta I=1/2$ rule for kaons.

At $O(p^2)$ in CHPT, $|\Delta| S=1$ amplitudes
can be described in terms of three couplings in octet symmetry,
\ba
{\cal L}_{\Delta S=1}^{(2)} &=&
-\frac{3 G_F}{5 \sqrt 2} V_{ud} V_{us}^* \, 
F_0^4 \left[ G_8 \, \langle u_\mu u^\mu \Delta_{32} \rangle
\right. \nonumber \\ &+&
 G_8' \, \langle \chi_{(+)} \Delta_{32} \rangle \nonumber \\
&+& \left. G_{27} \, t^{ij,kl} \langle u^\mu \Delta_{ij} \rangle  
\langle u_\mu \Delta_{kl} \rangle \right] +{\rm h.c.} \, 
\ea
We have pulled out the Fermi coupling constant, $G_F$, and
the relevant Cabibbo-Kobayashi-Maskawa matrix elements $V_{ij}$. 
 $U\equiv u u \equiv e^{i \sqrt 2 \Phi/F_0}$ with $\Phi$ a
SU(3) matrix  collecting the lowest pseudo-scalar meson
$\pi$, $K$, and $\eta_8$ fields;
$F_0$ is the chiral limit value of the pion decay constant
$f_\pi \simeq 92.4$ MeV;  
$D_\mu U$ is the covariant derivative acting on $U$ and 
$u_\mu \equiv  i u ^\dagger (D_\mu U) u $; 
$\chi_{(+)} \equiv  u^\dagger \chi u^\dagger
+ u \chi^\dagger u$ with $\chi \equiv  2 B_0 {\cal M}$, ${\cal M}$
is a 3 $\times$ 3 matrix collecting the light quark masses
and $ B_0 $ is proportional
to the quark condensate in the chiral limit,
$B_0 \equiv  - \langle 0 | \overline q q | 0 \rangle /F_0^2$.
The symbols $\Delta_{ij}$ and 27-plet tensor $t^{ij,kl}$  take into 
account for the correct flavour combinations and were defined 
in \cite{BPP98}.
At this order 
\ba
\left| \frac{A_0}{A_2} \right|^{(2)} = \sqrt 2 \, 
\left( \frac{9 G_8 + G_{27}}{10 G_{27}} \right) \, .
\ea
At leading order in $1/N_c$, $G_8=G_{27}=1$ and 
\ba
\left| \frac{A_0}{A_2} \right|^{(2)} = \sqrt 2 \, ; 
\ea
 i.e. more than
a factor ten lower that the experimental number !

\section{The Heavy $X_i$-Bosons Method: Matching Short-- and Long--Distances}

We analyse $|\Delta S|=1$ off--shell two-point Green functions
\ba
\Pi^{ij}(q^2) \hspace*{5cm} 
\nonumber \\ \equiv  i\int {\rm d}^4\, x \, e^{iqx}
\langle 0 | T\{ P^i(0)^\dagger 
P^j(x) \, e^{i \Gamma_{\Delta S=1}} \}| 0 \rangle  \nonumber \\
\ea
 in the presence of strong interactions.
These Green functions were studied in CHPT to $O(p^2)$ in \cite{BER85} and
to $O(p^4)$ in \cite{BPP98}. $P^i(x)$ are external pseudo-scalar sources that
couple to pion, kaon, and $\eta_8$ fields. The $\Delta S=1$ Standard Model
 effective action at some scale $\mu$ below the charm quark mass, 
can be written as
\ba
\Gamma_{\Delta S=1} \equiv -\frac{G_F}{\sqrt 2}  V_{ud} V_{us}^*
\, {\dis \sum_{i=1}} C_i (\mu) \, \int {\rm d^4} y \, Q_i(y)
\ea
with $C_i(\mu)$ Wilson coefficients which are known to two-loops and
$Q_i(y)$ are four--quark local operators inducing $\Delta S=1$ transitions.
The list of relevant  operators is given in \cite{BP99a}.

The effective action $\Gamma_{\Delta S=1}$ is generated by {\em virtual}
$W$--boson exchanges. This makes necessary the intervention of 
strong interactions at {\em all} scales between 0 and $\infty$
 to  calculate  weak matrix elements. Matching long-- and short--distances 
is the big challenge of calculating weak matrix elements. The procedure we
propose is to use an effective field theory which reproduces
the physics  of the four--quark  $\Gamma_{\Delta S=1}$  operators 
{\em below} some scale $\mu_L$  around the charm quark mass
through the exchange of heavy $X_i$--bosons.
For instance,
\ba
X_\mu^1 \left\{ g_1(\mu_L, M_X, \cdots) [\overline s_L \gamma^\mu d_L]
\right. \nonumber \\ + \left. 
g_1'(\mu_L, M_X, \cdots) [\overline u_L \gamma^\mu u_L] \right\}
\ea
reproduces the physics of $Q_1(x)$ {\em below} $\mu_L$. At $\mu_L$
we need matching conditions which fix $g_1(\mu_L, M_X, \cdots)$ 
and $g_1'(\mu_L, M_X, \cdots)$. 
A detailed example on how this procedure works 
 will be presented in \cite{BP99b}.

In this way, we resum large $\log(M_W/\mu) /N_c$ 
to any available order in the short--distance and
long--distance calculations. Scale and scheme dependence
are  correctly treated as well. 
 
In the heavy $X_i$--boson exchange effective field theory,  
the basic non--leptonic interaction for physics below $\mu_L$ is given by 
\ba
\sim g_1^\dagger \, g_1'
\int \frac{{\rm d}^4 r}{(2\pi)^4} \, \int e^{i q . x}
\frac{g_{\mu\nu}}{M_{X_1}^2} \, J^{\mu\dagger}_1(x) \, J^{\nu'}_1(0) \, .
\ea
Now, we can calculate analogously to what one does
for the $\gamma$--exchange contribution to 
 $\pi^+$--$\pi^0$ or $K^+$--$K^0$ \cite{BP97} mass difference.
We can separate the long-- and short--distance pieces using
an Euclidean cut--off $\mu$
\ba
\label{long}
\int {\rm d}^4 r_E \to \int {\rm d}\Omega \left[ {\dis \int^\mu_0}
{\rm d} |r_E| + {\dis \int^\infty_\mu} {\rm d} |r_E| \right] \, . 
\ea
The short--distance piece is then consistently treated within the
heavy $X_i$--boson exchange effective theory at next-to-leading order
in the $1/N_c$ expansion and the long-distance piece with an appropriate
hadronic model or data if available. In our case, 
the low--energy part is treated with the ENJL model presented
in \cite{ENJL}. At leading order in $1/N_c$, 
this  is a model with the same chiral structure as QCD and
which reproduces most of the low energy dynamics of the strong
interactions. These two features together with a reasonable matching
with short--distance QCD are expected to be the bulk of the dynamics 
needed to predict weak matrix elements 
and are the basis of our calculational method 
while lacking first principle calculations. 
Model dependence enters only through the evaluation
of the long--distance piece in (\ref{long}).

The ENJL model doesn't confine and does have a wrong high energy behaviour
at high energies. We smear out these bad features  by calculating
 far off--shell with very small momenta and using only fits
up to order five or six at most, see more details in \cite{BP99a}. 
There are good prospects to eliminate to a large extent the bad high 
energy behaviour
and enlarge beyond 1 GeV the matching between short-- and 
long--distances using the model in \cite{PPR98}.

\section{The $\Delta I=1/2$ Rule}
Here we give the main conclusions of our work.
Penguin--like diagrams with $Q_2$ dominate the octet coupling  
$G_8$ (around 63 \%) in the whole range
of scales studied  [between 0.5 GeV and 1. GeV] 
producing the observed huge enhancement. 
The penguin--operator $Q_6$ contribution to $G_8$ is around 12 \%.

\begin{figure}[thb]
\begin{center}
\vspace*{-0.7cm}
\leavevmode\epsfxsize=6.75cm\epsfbox{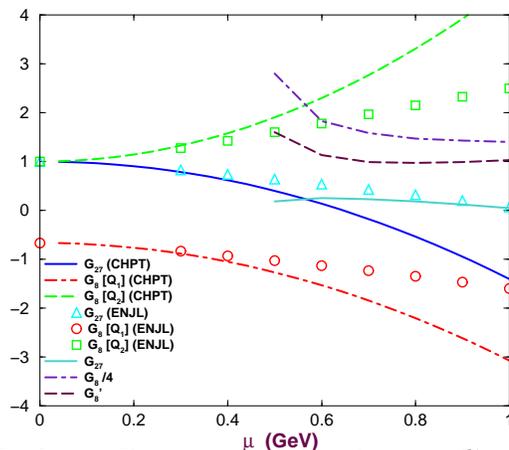}
\vspace*{-0.7cm}
\caption{\label{matching} We show the long--distance contributions to $G_8$, 
$G_{27}$, and $G_8'$ couplings using lowest order CHPT (quadratic dependence
in $\mu$) and ENJL. The final scheme independent 
result including short--distance at two-lops and matching is also shown.}
\vspace*{-0.7cm}
\end{center}
\end{figure}

\begin{figure}[thb]
\begin{center}
\vspace*{-0.7cm}
\leavevmode\epsfxsize=6.75cm\epsfbox{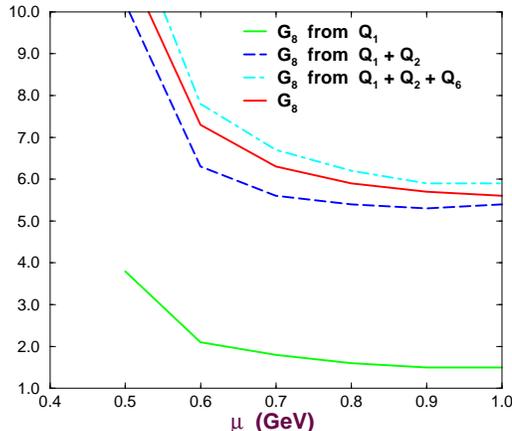}
\vspace*{-0.7cm}
\caption{\label{anatomy} The contributions of $Q_1$, $Q_1+Q_2$,
$Q_1+Q_2+Q_6$ to $G_8$. The final result for $G_8$ 
using short--distance at two--loops with the 
scheme dependence removed \cite{BP99a,BP99b} is also shown.}
\vspace*{-0.7cm}
\end{center}
\end{figure}

There is a large cancellation between 
the $B_K$--like diagrams contribution to $G_8$ from  $Q_1$ and $Q_2$.
 The relatively large positive contribution from $Q_1$  
is canceled by $B_K$--like diagrams from $Q_2$ to give in 
total less than 7 \% of $G_8$ from  $B_K$--like 
diagrams.  Factorizable contributions plus $B_K$--like contributions
are around 23 \% of $G_8$.
The sum of the rest of operators contributes by less than 5 \% and 
decreases $G_8$ up to its final value. 
More than 75 \%  of the value of $G_8$ comes from  penguin--like diagrams.
We show in Figure \ref{matching} 
the matching obtained for the three $O(p^2)$
couplings and in Figure \ref{anatomy}
 the relative contributions of $Q_1$, $Q_2$, and $Q_6$ to $G_8$.

At the same time, there are no penguin--like contributions to $G_{27}$
and $B_K$--like diagrams  for $Q_1$ and $Q_2$ decrease the 27--plet 
coupling from one to a value between one half and one third.
There is a large cancellation between the contributions of $Q_1$ and $Q_2$.
In summary, penguin--like diagrams with $Q_2$ dominate largely the enhancement 
of $G_8$ and $B_K$--like diagrams for  $Q_1 + Q_2$ produce  
the small value of $G_{27}$.
 These two facts are responsible for the $\Delta I=1/2$ rule
in (\ref{deltaI}). 

Experimentally
\ba
G_8 = 6.2 \pm 0.7 \, ; \hspace*{1cm} G_{27}= 0.48 \pm 0.06\, .
\ea
Here, we have only included the uncertainty
from the value of the pion decay constant in the chiral limit
$F_0=(86\pm10)$ MeV, 
since no uncertainties from the fit procedure were quoted in \cite{KMW91}.
We get
\ba
4.3 &<& G_8 < 7.5 \, ; \hspace*{1cm} 0.8 < G_8' < 1.1 \, ; \nonumber \\
0.25 &<& G_{27} < 0.40 
\ea
and 
\ba
15 < \left| \frac{A_0}{A_2} \right|^{(2)} < 40 .
\ea
This last result is somewhat large mainly because of the small value
we get for $G_{27}$. Notice that this calculation is to next-to-leading 
in $1/N_c$ and to {\em all} orders in Chiral Perturbation Theory. 
One can expect, therefore non--negligible $1/N_c^2$ corrections
--typically of the oder of (30$\sim$ 40) \%, but the main 
 $\Delta I=1/2$ enhancement is there. We want to stress that 
these are parameter free {\em predictions}, the three input values we  need
were fixed in  \cite{ENJL} from low energy phenomenology in the strong sector.
We believe there are good prospects
at obtaining {\em predictions} on $\Delta S=1$ transitions
and  $\varepsilon'/\varepsilon$ \cite{BPP99}.

\section{The $Q_6$ Penguin Operator}
In the chiral limit, the contribution of $Q_6$ to $G_8$ 
is proportional to the quark condensate squared. 
At leading order in $1/N_c$, the scale dependence of  $\langle \pi\pi | Q_6 
| K \rangle$ is  exactly canceled by the Wilson coefficient
$C_6(\mu)$ \cite{Q6}. We have shown in \cite{BP99a}
that the scale dependence is also canceled
 at next-to-leading in $1/N_c$ for the factorizable part.
Then, as for the rest of  $Q_i(y)$ operators,
the matching between short-- and long--distances becomes an affair
of non--factorizable contributions. 

Outside the chiral limit, the strange quark
 condensate squared {\em does not} factorize and most of its
 quark mass corrections produce actually the coupling $G_8'$
which does not contribute to $K\to \pi\pi$. 
Both $G_8$ and $G_8'$ are still proportional to the 
chiral limit value of the quark condensate squared, and 
  kaon and pion masses enter in higher order  CHPT corrections.
Therefore, the contribution to $G_8$ from $Q_6$ 
cannot be proportional to $1/m_s^2$  and the usual parameterization  
being inversely proportional to the strange quark mass squared is very 
misleading. In addition, the VSA result of 
$\langle (\pi\pi)_0 | Q_6 | K \rangle$ has an IR divergence 
as shown in \cite{BP99a}.
In view of all these problems, we propose to quote directly
values of matrix elements as we did in \cite{BP99a}. 
We give in Table \ref{tab:Q6} the results
for the contribution of $Q_6$ to $G_8$.
\begin{table}[thb]
\centering
\begin{tabular}{|c|c|c|}
\hline
Scale (GeV) & One-Loop & Two-loops (SI) \\ \hline 
 0.5 &0.98 & 2.14 \\
 0.6 &0.73 & 1.49 \\
 0.7 &0.53 & 1.10 \\
 0.8 &0.38 & 0.83 \\
 0.9 &0.26 & 0.62 \\
 1.0 &0.15 & 0.45 
 \\ \hline
\end{tabular}
\caption{\label{tab:Q6} The contribution of $Q_6$ to $G_8$
using short--distance to one--loop
and to two--loops with the the scheme dependence removed
see \cite{BP99a,BP99b}.} 
 \end{table}

However, for the sake of comparison with other results in the literature
which only quote $B_6$--parameters, we give our results for
\ba
B_6^{(1/2)}( \mu) &\equiv & \langle (\pi\pi)_0 | Q_6 | K \rangle
/ \left[\langle (\pi\pi)_0 | Q_6 | K \rangle {\Large|}_{N_c}\right] 
\nonumber  \\ 
\ea
where
\ba
\label{B6Nc}
\langle (\pi\pi)_0 | Q_6 | K \rangle {\Large|}_{N_c} 
= -i \, 32 \, \frac{G_F}{\sqrt 2} \, V_{ud} V_{us}^* \, C_6(\mu)
\nonumber \\
\times \, F_0 (m_K^2-m_\pi^2) \,  
\frac{\langle 0 | \overline q q | 0 \rangle^2 (\mu)}{F_0^6}
L_5(\nu) \, . \nonumber \\
\ea
$\langle 0 | \overline q q | 0 \rangle$ is the quark condensate
in the chiral limit which in a very good approximation we take  
to be the average of up and down quark condensate  \cite{BPR95}. 
With this definition, we avoid the IR  divergence in the VSA value
of $\langle (\pi\pi)_0 | Q_6 | K \rangle$ \cite{BP99a}. 
The large $N_c$ result (\ref{B6Nc}) contains still the ambiguity 
in the value of the scale $\nu$  which is only canceled
by the IR divergent part. We fix $\nu=M_\rho$
 and use $L_5(M_\rho)=(1.4\pm0.3) \, 10^{-3}$.

To lowest CHPT order 
$p^2$ and next-to-leading in $1/N_c$, we get
\ba
B_6^{(1/2)}(\mu) = 0.76 \pm 0.20
\ea
 in agreement with \cite{HKPS99}. 
To this order the scale dependence $\mu$ 
is canceled exactly due to the need of
canceling the IR divergence. Notice also
that the value of $B_6^{(1/2)}$ very near to one  
is due to the large  cancellation between the 
two types of factorizable contributions \cite{BP99a}. This cancellation
is  exact at order $p^0$
and very large at order $p^2$ due to the cancellation of the IR divergence.
 It does not however protect the value of $B_6^{(1/2)}$ from
higher CHPT order corrections.
In fact, also in the chiral limit but 
to {\em all} orders in CHPT and next-to-leading in $1/N_c$, we get
\ba
B_6^{(1/2)}( \mu) = 2.2 \pm 0.5
\ea
 for scales $\mu \in [0.6, 1.0]$ GeV. The scale dependence is very
mild. The importance of higher order  CHPT corrections is manifest in 
this result. A large enhancement of $B_6^{(1/2)}$ was also obtained
in \cite{HKPS99} when $O(p^4)$ corrections were included.

We found that the $G_8$ enhancement 
is not due to the contribution of the penguin operator $Q_6$. 
 The only relation between the dynamics underlying
the value of $\varepsilon'/ \varepsilon$ in the SM and the 
large value of $G_8$ is  the type of dominant diagrams, namely, 
penguin--like diagrams. But, $\varepsilon'/ \varepsilon$  is dominated by
the penguin operators
$Q_6$ and $Q_8$ while $G_8$  by the $Q_2$ operator. 

This work has been done in an enjoyable collaboration with Hans Bijnens.
It is a pleasure to thank Stephan Narison for the invitation
to this very interesting conference.

\end{document}